\documentstyle[aps,prd,eqsecnum,twocolumn,epsf]{revtex}

\newcommand{\square}{\kern1pt\vbox{\hrule height 1.2pt\hbox{\vrule width 
1.2pt\hskip 3pt\vbox{\vskip 6pt}\hskip 3pt\vrule width 0.6pt}\hrule height
0.6pt}\kern1pt}
\newcommand{\beq}{\begin{equation}}
\newcommand{\beqn}{\begin{eqnarray}}
\newcommand{\eeq}{\end{equation}}
\newcommand{\eeqn}{\end{eqnarray}}

\begin{document}

\draft

\twocolumn[\hsize\textwidth\columnwidth\hsize\csname@twocolumnfalse\endcsname

\title{Five-dimensional Black Hole and Particle Solution with Non-Abelian  Gauge
Field}

\author{Naoya Okuyama$^1$ and Kei-ichi Maeda$^{1,2,3}$,}

\address{$^1$Department of Physics, Waseda University, Okubo 3-4-1, Shinjuku, 
Tokyo 169-8555, Japan\\[-1em]~}
\address{$^2$ Advanced Research Institute for Science and Engineering, Waseda 
University, Shinjuku, Tokyo 169-8555, Japan\\[-1em]~}
\address{$^3$ Waseda Institute for Astrophysics, Waseda University, Shinjuku, 
Tokyo 169-8555, Japan\\[-1em]~}

\date{\today}

\maketitle

\begin{abstract}
We study the 5-dimensional Einstein-Yang-Mills system with a cosmological 
constant. Assuming a spherically symmetric spacetime, we find a new analytic
black hole solution,  which approaches asymptotically ``quasi-Minkowski", ``quasi
anti-de Sitter", or  ``quasi de Sitter" spacetime depending on the sign of a
cosmological constant. Since there is no singularity except for the origin which
is covered by an event horizon, we regard it as a localized object. This solution
corresponds to a magnetically charged  black hole.
 We also present a singularity-free particle-like 
solution and a non-trivial black hole solution numerically.
Those solutions correspond to the Bartnik-McKinnon solution and a colored black 
hole  with a cosmological
constant in the 4-dimensions. We analyze their asymptotic behaviors, spacetime
structures  and thermodynamical properties.
We show that  there is a set of stable solutions  if a cosmological constant  is
negative. 
\end{abstract}
\vskip 1pc
]

\section{Introduction}
 
Recent progress in a superstring theory shows that different string
theories are connected with each other via dualities, making them to be
unified to the M theory in 11 dimensions\cite{M-theory}. 
This provides us a motivation to study a higher
dimensional gravitational theory.
 String theory also predicts a boundary layer, a {\em brane},
 on which  edges of open  strings stand\cite{brane}. 
This  suggests a new perspective in cosmology, that is,  we are
living in a brane world, which is a three-dimensional hypersurface in a
higher-dimensional spacetime. 
 In contrast to the already familiar Kaluza-Klein picture 
in which  we live in 4-dimensional spacetime with $n$ dimensional
compactified ``internal space", 
our world view appears to be changed completely.  
Particles in the standard model are expected to be confined to the brane,
 whereas the gravitons propagate in the
entire bulk spacetime.

In the brane world cosmological scenario\cite{brane_cosmology},
 a higher-dimensional black hole solution
plays an important role. 
 Our universe is just a domain wall expanding in the
black hole background spacetime\cite{Kraus}.  The black hole mass gives a 
contribution  to dark radiation through its tidal force.
Hence, a higher-dimensional black hole or a globally regular solution 
with a cosmological constant is now a very interesting subject. 
In particular, in the context of the AdS/CFT
correspondence\cite{maldacena} or proposed dS/CFT 
correspondence\cite{strominger}, 
since the 5-dimensional Einstein gravity with a cosmological constant
gives a description of 4-dimensional conformal field theory in large $N$
limit, many authors study such localized objects in 5-dimensions\cite{5DBH}.

However, from a view point of brane cosmology, 
a black hole solution has a singularity in a bulk spacetime, although it is
covered by a horizon.  If a string theory or M theory is fundamental,
such a singularity should not exist. Then, if we can construct some
non-singular object in the bulk spacetime, it might be a manifestation
of singularity avoidance immanent in a fundamental theory.
In 4-dimensions, Bartnik and McKinnon found a particle-like solution as a 
globally  regular spacetime in a spherically symmetric Einstein-Yang-Mills  system
with SU(2) gauge group\cite{bartnik}. Soon after, a colored
black hole solution with a nontrivial non-Abelian structure was also
found\cite{bizon}.  These  solutions were also extended to those in the system
with a cosmological constant\cite{volkov,torii,bjoraker}. 
From stability analysis, it turns out that the solution with   zero 
or positive cosmological constant is unstable\cite{brodbeck}, while those 
 with negative
cosmological constant is stable\cite{bjoraker,winstanley}.
Since a negative cosmological constant is naturally expected in a brane world
scenario just  as the Randall-Sundrum  model\cite{RS}, the above fact is very
interesting. In this paper, then, we study a nontrivial
particle-like solution or black hole solution in 5-dimensions with a 
cosmological constant. 

As for non-Abelian gauge field in a bulk spacetime,
although gauge interactions are confined on a brane and 
Yang-Mills fields are expected to exist only in the brane,
if our 5-dimensional spacetime is obtained as an effective theory, 
it may not be the case.
In fact, Lukas, et al. \cite{Lukas} showed that U(1)-field appears in the effective 
5-dimensional bulk spacetime, from dimensional reduction of the Ho\v{r}ava-Witten 
model\cite{M-theory}.
We may find non-Abelian gauge field from some other type dimensional reduction
of a unified theory.

There is another interesting point to discuss non-Abelian gauge fields in a bulk.
Using a brane structure, new mechanism of spontaneous symmetry breaking of
gauge interactions have been proposed\cite{hosotani}.
In this picture, the present standard model (SU(3)$\times$SU(2)$\times$U(1)) is
obtained on the brane assuming some higher-symmetric gauge interactions such as 
SU(5) in the bulk.

Therefore, in this paper, we assume that non-Abelian gauge field
appears in 5-dimensional bulk spacetime.
In \S.\ref{sec:basicequations}, we first  derive the basic equations of a
spherically symmetric  Einstein-Yang-Mills system in 5-dimensions.
With a spherically symmetric ansatz, the gauge potential of SU(2) Yang-Mills 
field is be decomposed into the ``electric" and   ``magnetic" part, which
derivation is given in Appendix \ref{sec:appendix}. 
There is a nontrivial analytic solution in the case with ``magnetic" field,
which corresponds to a magnetically charged black hole in 4-dimensions.
This analytic solution and its property are examined in 
\S.\ref{sec:analyticsolutions}. We also present non-trivial particle-like and 
black hole 
 solutions, which correspond to the Bartnik-McKinnon type and colored black hole
type  solutions in 4-dimensions, in \S.\ref{sec:numericalsolutions}. 
We also analyze those stability  in \S.\ref{sec:stability}. 
Summary and discussion follow in 
\S.\ref{sec:summaryanddiscussion}.

\section{Basic Equation}
\label{sec:basicequations}

In order to find a black hole and particle-like solution of the 5-dimensional 
Einstein-Yang-Mills system, we first write down the basic equations. The action
is given by
\begin{eqnarray}
S=\frac{1}{16\pi}\int d^5x\sqrt{-g_5}\left[{1\over G_5}(R-2\Lambda) -{1\over
g^2}{\rm Tr}{\bf F}^2\right],
\end{eqnarray}
where $G_5$ is a 5-dimensional gravitational constant, $\Lambda$ is a 
5-dimensional cosmological constant, and $g$ is a gauge coupling constant. Now we
adopt gauge group is SU(2).
${\bf F}=F_{\mu\nu}dx^\mu\wedge dx^\nu$ is a field strength of the gauge field, 
which is described by the vector potential ${\bf A}=A_\mu dx^\mu$ as
\begin{eqnarray}
F_{\mu\nu}=\partial_\mu A_\nu-\partial_\nu A_\mu - [A_\mu,A_\nu].
\end{eqnarray}
Defining the 5-dimensional Planck mass by $m_5=G_5^{-1/3}$ and a fundamental mass 
scale of the gauge field by
$m_g=g^{-2}$, we introduce a typical length scale of the present system, which is 
given by 
\begin{eqnarray}
\lambda=\left({m_g\over m_5^3}\right)^{1\over 2}=\left({G_5\over g^2}\right)^{1\over 2}.
\end{eqnarray}
We will normalize a scale
length by this $\lambda$.

We consider a spherically symmetric 5-dimensional spacetime, which metric is 
given by
\begin{eqnarray}
ds^2=\lambda^2\left[-f(t,r) e^{-2\delta(t,r)}dt^2+\frac{dr^2}{f(t,r)} +r^2 
d\Omega^2_3\right],\label{metric}
\end{eqnarray}
where
\begin{eqnarray}
f(t,r)&=&1-\frac{\mu(t,r)}{r^2}+\epsilon \frac{r^2}{\ell^2},\\
d\Omega^2_3&=&d\psi ^2 + \sin ^2\psi \left(d\theta ^2+\sin^2 \theta ~d\varphi ^2 
\right),
\end{eqnarray}
where we set $\Lambda=-6\epsilon/(\lambda\ell)^2$ with $\epsilon=0$ or $\pm 1$, 
corresponding to the signature of $\Lambda$, i.e. $\epsilon=1,0,$ and $-1$
corresponds to $\Lambda<0, \Lambda=0$, and  $\Lambda>0$, respectively. Note that
$t, r, $and $\mu$ are all dimensionless variables. We shall call $\mu$ a `mass'
function.
$\ell$ denotes the ratio of the length scale of a cosmological constant to 
$\lambda$.

From Appendix A, we find a generic form of a spherically symmetric SU(2) gauge 
potential. If we take only an ``electric" part of the field, the gauge potential
is given by Eq.~(\ref{pot-ele}), which yields  the basic equations as
\begin{eqnarray}
&&\mu'=\frac{2}{3}r^3(A'e^\delta)^2,\\
&&\dot{\mu}=0,\\
&&\delta'=0,\\
&&\left[\left(A'e^\delta\right)^2 \right]^\cdot  =0,\\
&&\left[\left(A'e^\delta\right)^2\right]'+\frac{6}{r}\left(A'e^\delta\right)^2=0,
\end{eqnarray}
where a dash and dot denote the partial derivative with respect to $r$ and $t$ 
respectively. This equation gives the Reissner-Nordstrom type solution such as
\begin{eqnarray}
\mu&=&{\cal M}-\frac{2Q^2}{3r^2},\\
\delta&=&0,\\
A&=&-\frac{Q}{r^2}.
\end{eqnarray}
This result is same to the case of 4-dimensions.

If the ``magnetic" part of the gauge field, which is given by 
Eq.~(\ref{pot-mag}), appears, we find another basic equations as follows. Using
the gauge freedom, we set $X=0$, resulting in the gauge potentials as
\begin{eqnarray}
A_t^a&=&0, ~~~ A_r^a=0,\\
A_\psi^a&=&(0, 0, w),\\
A_\theta^a&=&(w\sin\psi, -\cos\psi, 0),\\
A_\varphi^a&=&(\cos\psi\sin\theta, w\sin\psi\sin\theta, -\cos\theta),
\end{eqnarray}
where we set  $\phi=w(r,t)$.
With the above ansatz, we find the Einstein equations and Yang-Mills equation of 
the present system as
\begin{eqnarray}
\mu'&=&2r\left[fw'^2+f^{-1}e^{2\delta}\dot{w}^2 +\frac{(1-w^2)^2}{r^2}\right],
\label{equationofm}\\
\dot{\mu}&=&4 rfw'\dot{w},\label{equationofm2}\\
\delta'&=&-\frac{2}{r}[w'^2+f^{-2}e^{2\delta}\dot{w}^2]\label{equationofdelta},
\end{eqnarray}
and
\begin{eqnarray}
\frac{1}{r}(rfe^{-\delta}w')'+\frac{2}{r^2}e^{-\delta}w(1-w^2)=\left(f^{-1}
e^{\delta}\dot{w}\right)^\cdot.\label{equationofw}
\end{eqnarray}
Eqs.~(\ref{equationofm})-(\ref{equationofw}) look very similar to those in  the 
case of the 4-dimensional Einstein-Yang-Mills system. However, a little
difference of the power exponent of $r$ brings a big difference in the behavior
of solutions, as we will see later.

\section{Analytic solutions}\label{sec:analyticsolutions}

Now we look for a ``magnetic" type static solution of the system 
(\ref{equationofm})-(\ref{equationofw}). Dropping the time derivative terms, we
find the basic equations as
\begin{eqnarray}
&&\mu'=2r\left[fw'^2+\frac{(1-w^2)^2}{r^2}\right],\label{eq_m}\\
&&\delta'=-\frac{2}{r}w'^2,\label{eq_delta}\\
&&\frac{1}{r}(rfe^{-\delta}w')'+\frac{2}{r^2}e^{-\delta}w(1-w^2)=0.\label{eq_w}
\end{eqnarray}

The above differential equations (\ref{eq_m})-(\ref{eq_w}) have two analytic 
solutions. One analytic solution is
\begin{eqnarray}
w=\pm 1,~~\mu={\cal M}, ~~\delta = 0,\label{schwarzschildsol}
\end{eqnarray}
which corresponds to the Schwarzs\-child or the Schwarzs\-child-anti de Sitter 
(or de Sitter) spacetime, which properties are well known.

Another analytic solution is given by
\begin{eqnarray}
w=0,~~\mu={\cal M}+2\ln r, ~~\delta = 0.
\label{nontrivialsol}
\end{eqnarray}
This solution has a nontrivial geometry.
In the 4-dimensional spacetime, this type of solution describes the 
Reissner-Nordstrom type geometry with a magnetic charge. In the
5-dimensional spacetime, $2\ln r$ term appear in the mass function $\mu$.
Although $\mu$ diverges, the metric itself approaches that of well-known
symmetric spacetime for each $\epsilon$, i.e. the Minkowski, de Sitter and anti
de Sitter one. We first study the properties of this solution in the following
subsections.

\subsection{Asymptotic structure}

Since the mass function diverges, we have to  analyze carefully those  asymptotic
behaviors. For the case of $\epsilon=0$, the Riemann curvature is finite except
at $r=0$ and vanishes at infinity as
\begin{eqnarray}
R_{\mu\nu\rho\sigma}R^{\mu\nu\rho\sigma}\rightarrow 288 \left({\ln r\over
r^4}\right)^2 .
\end{eqnarray}
For the case of $\epsilon=\pm 1$, the Riemann curvature is also finite 
everywhere except at 
$r=0$ and converges as
\begin{eqnarray}
R_{\mu\nu\rho\sigma}R^{\mu\nu\rho\sigma}\rightarrow {40 \over 
\ell^4}
\end{eqnarray}
as $r\rightarrow \infty$. This finite value just comes from
the Ricci curvature. 
The metric form approaches
\begin{eqnarray}
f(r)\rightarrow 1+\epsilon~{r^2\over \ell^2}
\end{eqnarray}
as $r\rightarrow \infty$.
These spherically symmetric and static spacetimes are singular only at $r=0$, 
and seem to approach a ``maximally symmetric spacetime". Therefore we may
recognize it as an localized object in such a ``maximally symmetric spacetime".

However we have to analyze those asymptotic behaviors more carefully.
The asymptotically flatness condition is mathematically defined  using the 
conformal  transformation. We can also extend this formulation for an
asymptotically de Sitter (or anti-de Sitter) spacetime as well as for a
higher-dimensional spacetime.

In an asymptotically flat spacetime, we can naturally define a mass of an 
isolated object, which is called the ADM mass \cite{adm}.
It is defined by
\begin{eqnarray}
G_5 M_{\rm ADM} ={1\over 16\pi}\oint_{I_0} dS_i\left[\partial_j h^{ij}
-\eta^{ij}\partial_j h^k_k\right]
\end{eqnarray}
in 5-dimensional spacetime, where $\eta_{\mu\nu}$ is Minkowski metric  and 
$h_{\mu\nu}=g_{\mu\nu}-\eta_{\mu\nu}$.
$dS_i$ is an infinitesimal surface element of spacelike infinity $I_0$.
For the present non-trivial solution with $\epsilon =0$, we find
\begin{eqnarray}
G_5 M_{\rm ADM} =\lim_{r\rightarrow \infty}{3\pi \over 8}\lambda^2 \left({\cal M}+2\ln 
r\right)
\end{eqnarray}
which diverges as $\ln r$.
The coefficient $3\pi/8$ appears just because Eq.~(\ref{equationofm}) yields  
\begin{eqnarray}
\mu={3\pi\over 8}~\int dv~[-T^0_{~0}]. 
\end{eqnarray}
For $\epsilon =-1$, if the spacetime
is asymptotically de Sitter, we can also introduce a conserved mass, which is
called the Abbott-Deser mass defined by \cite{ad,nakao}.
If the spacetime is asymptotically de Sitter, 
$M_{\rm AD} =M_{\rm ADM}$,
which diverges again as  $\ln r$.

In a 5-dimensional asymptotically anti-de Sitter spacetime, we can also  define a
conserved mass associated with a timelike Killing vector $\vec{\xi}$ at the
3-sphere  $C$ on conformal infinity  ${\cal I}$ as \cite{ashtekar,ashtekar2}
\begin{eqnarray}
G_5 M_{\xi}[C]:=-\frac{\lambda^2\ell}{16\pi}\oint_C{\cal E}_{\mu\nu}\xi^\mu 
dS^\nu,
\end{eqnarray}
where ${\cal E}_{\mu\nu}$ is the electric part of Weyl tensor defined by
\begin{eqnarray}
{\cal E}_{\mu\nu}:= {\ell^2\over \Omega^{2}}C_{\mu\rho\nu\sigma}n^\rho 
n^\sigma.
\end{eqnarray}
$\Omega$ is a conformal factor and $n_\mu=\nabla_\mu \Omega$.
In the case of Schwarzschild-anti de Sitter spacetime (\ref{schwarzschildsol}), 
this mass gives ${\cal M}$. In the non-trivial solution, however, this quantity is
calculated on the 3-sphere $C$  with a radius $r$ as
\begin{eqnarray}
G_5 M_{\xi}[C]= {3\pi\over 8} \lambda^2\left[{\cal M}+2\ln r -{7\over
6}\right].
\end{eqnarray}
It diverges as $\ln r$ as $r\rightarrow \infty$.

In any case, the ``mass" is not finite, which means that ``total energy" of the
system is not finite. Therefore, strictly speaking, 
 we  should not regard it as an isolated object.
However, there is no singularity except at $r=0$ and the metric form itself 
approaches either Minkowski or de-Sitter (anti-de Sitter) one. Hence, we call it
a ``quasi-isolated" object. We remind that we know a similar ``isolated" object,
i.e. a 4-dimensional self-gravitating global monopole. Its metric is described as
\begin{eqnarray}
ds^2=-f(r)dt^2+f(r)^{-1}dr^2 +r^2d\Omega^2,
\end{eqnarray}
where $f(r) \equiv 1-{2m(r)/ r} \sim 1-\alpha -2M/r +O(1/r^2)$.
In this case, the mass function $m(r) $ diverges as $M+\alpha r/2 +O(1/r)$ as 
$r\rightarrow \infty$. In fact the ADM mass diverges.
Rescaling the time and radial coordinates as $r\rightarrow (1-\alpha)^{1/2} r$ 
and $t\rightarrow (1-\alpha)^{-1/2} t$, we can rewrite the metric form as
\begin{eqnarray}
ds^2=-f(r)dt^2+f(r)^{-1}dr^2 +(1-\alpha) r^2d\Omega^2,
\end{eqnarray}
where $f(r)=1-2\tilde{M}/r$ with $\tilde{M}=M(1-\alpha)^{-3/2}$.
This spacetime looks asymptotically flat but has a deficit angle $\alpha$.
 Nucamendi and Sudarsky showed that this spacetime is asymptotically 
simple but not asymptotically empty\cite{nucamendi}. They called it a
``quasi-asymptotically flat" spacetime and defined a new mass for a spacetime
with a deficit angle, which is a generalization of the ADM mass,
using the first law of black hole thermodynamics.

In our case, the mass function diverges as $\ln r$, which is less divergent than 
the case with a deficit angle ($r^2$ in 5-dimensions). Then we can also call such
a spacetime ``quasi-asymptotically" flat or ``quasi-asymptotically" de Sitter
(anti-de Sitter) one.

\subsection{Spacetime structure: horizon and singularity}

This solution has a horizon, where
\begin{eqnarray}
f(r)=1-{{\cal M}+2\ln r\over r^2}+\epsilon {r^2\over \ell^2}=0.
\label{eq:horizon}
\end{eqnarray}
We study those horizons and the singularity separately for each value of 
$\epsilon$.

\subsubsection{$\epsilon=0$}

In this case, if ${\cal M}>1$, Eq.~(\ref{eq:horizon}) has two roots $r=r_\pm 
(r_-<r_+)$, which correspond to two horizons, $r_+$ corresponds to an event
horizon, while $r_-$ is an inner horizon. A timelike singularity appears at
$r=0$. For the case of ${\cal M}=1$, two horizons degenerate and a black hole 
becomes extreme. If
${\cal M}<1$ there is no horizon, so a naked singularity appears.

\subsubsection{$\epsilon=1$}

This case also has two horizons $r=r_\pm (r_-<r_+)$ if ${\cal M}>  {\cal M}_{\rm
cr}$.
$r_+$ and  $r_-$ are an event horizon and an inner horizon, respectively.
The critical mass parameter ${\cal M}_{\rm cr}$ is given by the horizon radius 
of the extreme case ($r_{+ {\rm cr}}$), i.e.
\begin{eqnarray}
{\cal M}_{\rm cr}={1\over 2}\left(1+r_{+ {\rm cr}}^2\right)-2\ln r_{+ {\rm cr}},
\end{eqnarray}
where
\begin{eqnarray}
r_{+ {\rm cr}}={\ell\over 2}\left(-1+\sqrt{1+{8\over \ell^2}}\right)^{1\over 2}.
\end{eqnarray}
${\cal M}_{\rm cr}$ is always larger than unity and it approaches 1 as  $\ell
\rightarrow \infty$, which corresponds to the case of $\epsilon =0$. A timelike
singularity appears at $r=0$.
For the case of ${\cal M}={\cal M}_{\rm cr}$, the black hole is extreme, and
for${\cal M}<{\cal M}_{\rm cr}$, a horizon disappears.

\subsubsection{$\epsilon=-1$}

If a cosmological constant is positive, we expect a cosmological horizon  just as
a de Sitter spacetime. In fact, we always find at least one horizon.
If $\ell>2\sqrt{2}$, and
\begin{eqnarray}
{\cal M}_{\rm min}<{\cal M}< {\cal M}_{\rm max},
\end{eqnarray}
where ${\cal M}_{\rm min}=g(r_{+{\rm cr}})$ and ${\cal M}_{\rm max} =g(r_{-{\rm
cr}})$ with $g(r)=r^2-2\ln r -r^4/\ell^2$ and
\begin{eqnarray}
r_{\pm {\rm cr}}={\ell\over 2}\left(1 \pm\sqrt{1-{8\over \ell^2}}\right)^{1\over
2},
\end{eqnarray}
we find three horizons, $r_- (< r_{- {\rm cr}}) <r_+ (< r_{+ {\rm cr}}) <r_c$.
$r_-$, $r_+$ and $r_c$ are an inner,  event, and
cosmological  horizon, respectively. When ${\cal M}={\cal M}_{\rm max}$, the inner
and event horizons degenerate ($r_- =r_+$), while if  ${\cal M}={\cal M}_{\rm
min}$, the  event and  cosmological horizons coincide ($r_+ =r_c$). In the limit
of
$\ell\rightarrow 2\sqrt{2}$, ${\cal M}_{\rm min}= {\cal M}_{\rm max}= {\cal
M}_{\rm cr}=3/2 -\ln 2\approx 0.80685$, and then three horizons degenerate for
${\cal M}= {\cal M}_{\rm cr}$. 

For other cases, we have only one  horizon. The singularity at $r=0$ 
becomes naked.

We summarize the type of horizons in Table 1.

\vskip 1cm
\begin{tabular}{|l|l|c|}
\hline
 & $~{\cal M}<1$ &0\\
\cline{2-3}
$~\epsilon=0$ & $~{\cal M}=1$ &D\\
\cline{2-3}
 & $~{\cal M}>1$ & I, E\\
\hline
 & $~{\cal M}<{\cal M}_{{\rm cr}}$ &0\\
\cline{2-3}
$~\epsilon=1$ & $~{\cal M}={\cal M}_{{\rm cr}}$ & D\\
\cline{2-3}
 & $~{\cal M}_{{\rm cr}}<{\cal M}$ & I, E\\
\hline
 & 
$~{\cal M}<{\cal M}_{{\rm min}}$ & I\\
\cline{2-3}
 & 
$~{\cal M}={\cal M}_{{\rm min}}$ & I, D\\
\cline{2-3}
$~\epsilon=-1$, $\ell>2\sqrt{2}~$ & 
$~{\cal M}_{{\rm min}}<{\cal M}<{\cal M}_{{\rm max}}$ &~I, E, C\\
\cline{2-3}
 & 
$~{\cal M}={\cal M}_{{\rm max}}$ & D, C\\
\cline{2-3}
 & $~{\cal M}> {\cal
M}_{{\rm max}}~$ &C\\
\hline
$~\epsilon=-1$, $\ell\leq 2\sqrt{2}$ & & C\\
\hline
\end{tabular}
~\\
\vspace{.2cm}

\noindent
Table 1 : Type of horizons. I, E, C and D denote an inner,
event, cosmological and degenerated horizon, respectively. ``0" means no
horizon.
${\cal M}_{{\rm cr}},
{\cal M}_{{\rm min}}$, and 
${\cal M}_{{\rm max}}$ are defined in the text.

\subsection{Thermodynamical properties}

Next we shall see thermodynamical properties.
The Hawking temperature is easily calculated from a regularity condition at the 
event horizon\cite{hawking}. We find
\begin{eqnarray}
T_{BH}={1\over 2\pi r_+}\left[1-{1\over r_+^2}+2\epsilon 
{r_+^2\over\ell^2}\right].\label{temperature}
\end{eqnarray}
The entropy $S=A/4$ is given by
\begin{eqnarray}
S={1\over 2}\pi^2r_+^3,
\label{entropy}
\end{eqnarray}
because the volume of a unit 3-sphere is $2\pi^2$. 
Since the solution does not satisfy the asymptotically flat or de Sitter (or anti
de Sitter) conditions, we cannot define gravitational mass.
However, if we use the first law of thermodynamics
just as the case of a global monopole with a deficit angle\cite{nucamendi},  we
can define thermodynamical mass $M_T$ as $d M_T=TdS+\Phi dQ$. We find
\begin{eqnarray}
M_T={3\pi \over 8}{\cal M},
\label{mass_parameter2}
\end{eqnarray}
where an integration constant is set to be zero.
This result shows that the mass parameter ${\cal M}$ 
essentially denotes the thermodynamical mass.

\begin{figure}[tbh]
\epsfxsize = 2.1in
\epsffile{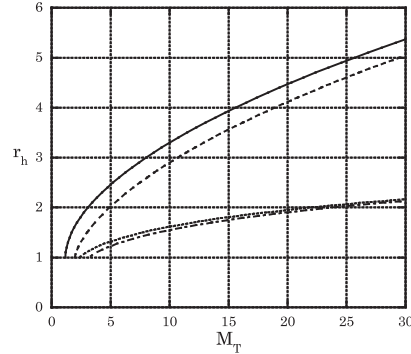}
\caption{$M_T$-$r_+$ relation. The horizon radius $r_+$ is depicted in terms of
thermodynamical mass $M_T$  for
$\epsilon=0$ and $\epsilon = 1$ by the solid and dotted line respectively.
That for the Reissner-Nordstrom
solution with same charge for $\epsilon=0$ and $\epsilon = 1$ is given by the dash and dotted-dash line as reference, respectively.}
\label{fig_mrh_analytic}
\end{figure}

\begin{figure}[tbh]
\epsfxsize = 2.1in
\epsffile{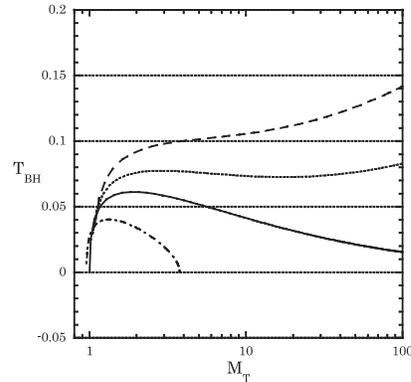}
\caption{$M_T$-$T_{BH}$ relation.
The solid line depicts the relation for $\epsilon=0$, while the dotted and dash
 lines represent those for $\epsilon=1$ with $\ell= 6.0$ and $\ell=4.0$,
and the dotted-dash line correspond to that  for $\epsilon=-1$ with $\ell=5.0$
respectively.}
\label{fig_mt}
\end{figure}

From Eq. (\ref{eq:horizon}), the  thermodynamical mass is given by
the horizon radius as
\begin{eqnarray}
M_T={3\pi \over 8}\left[r_+^2\left(1+\epsilon {r_+^2\over \ell^2}\right)
-2\ln r_+\right] .
\label{thermodynamical_mass}
\end{eqnarray}
We depict the $M_T$-$r_+$ relation in Fig. \ref{fig_mrh_analytic}. 
We find that the horizon radius is smaller than that of the
electrically charged Reissner-Nordstrom black hole.
We also show the  $M_T$-$T_{BH}$ relation  in
Fig.~\ref{fig_mt}. 
From Eqs.~(\ref{temperature})
and (\ref{mass_parameter2}), we find
\begin{eqnarray}
{dT_{BH}\over dM_T}=-{2\over 3\pi^2 r_+^3}{1-3/ r_+^2-2\epsilon r_+^2/\ell^2 
\over 1 - 1/r_+^2 + 2\epsilon r_+^2/\ell^2},
\end{eqnarray}
which gives a turning point where a specific heat changes its sign.
For the case of $\epsilon=0$, a specific heat is positive in $1<r_+<\sqrt{3}$ but 
becomes negative for $r_+ >\sqrt{3}$. 
(The corresponding critical value for thermodynamical mass is obtained by Eq.
(\ref{thermodynamical_mass}).) 
For the case of $\epsilon=1$, if
$\ell
\leq 2\sqrt{6}$, a specific heat is always positive. If $\ell > 2\sqrt{6}$, a
specific heat is positive in
$r_{+{\rm cr}}<r_+<r_{-{\rm ch}}$ and in $r_+>r_{+{\rm ch}}$, while it is
negative in $r_{-{\rm ch}}<r_+<r_{+{\rm ch}}$, where
\begin{eqnarray}
r_{\pm{\rm ch}}={\ell \over 2}\left(1\pm \sqrt{1-{24\over \ell^2}}\right)^{1\over 
2}.
\end{eqnarray}
For the case of $\epsilon=-1$, a specific heat is positive in $r_{-{\rm cr}}< r_+ 
< r_{\rm ch}$, while it is negative for $r_{\rm ch} < r_+ \leq r_{+{\rm cr}}$,
where
\begin{eqnarray}
r_{\rm ch}={\ell \over 2}\left(-1+\sqrt{1+{24\over \ell^2}}\right)^{1\over 2}.
\end{eqnarray}

\section{Numerical solutions}
\label{sec:numericalsolutions}

Just as 4-dimensions 
\cite{bartnik,bizon,bjoraker,winstanley}, we can find non-trivial structure of
self-gravitating Yang-Mills field. We obtain those solutions numerically.
We discuss two
cases; a particle solution and a black hole, separately. 
Here, we analyze only the case of $\epsilon=0$ or $1$.

\subsection{Particle solution}

In the case of a particle solution, we have to impose  regularity at the origin 
$r=0$. Since Eqs.~(\ref{equationofm})-(\ref{equationofw}) are invariant under the
transformation of $w\rightarrow -w$, we can set $w(0)>0$ without loss of
generality. Expanding $\mu$ and $w$ around $r=0$, we find those behaviors near
the origin as
\begin{eqnarray}
\mu(r)&=&4b^2 r^4+O(r^5)\\
\delta(r)&=&-4b^2r^2+{4\over 3}b^2\left({4\epsilon \over
3\ell^2}-3b-8b^2
\right)r^4 +O(r^5)\\ w(r)&=&1+br^2-{b\over 6}\left({4\epsilon \over
3\ell^2}-3b-8b^2
\right) r^4 +O(r^5)
\end{eqnarray}
with one free parameter $b$.
Using this boundary condition, we integrate the basic equations by the 
Runge-Kutta method.

\begin{figure}[tbh]
\epsfxsize = 2.1in
\epsffile{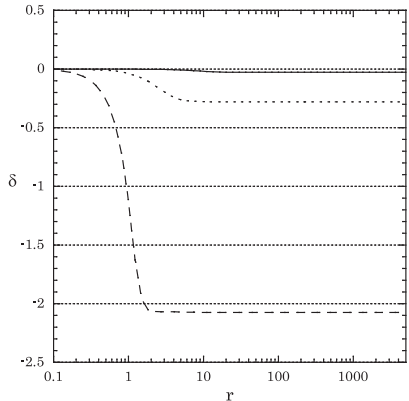}
\caption{The metric function $\delta(r)$ for a particle-like solution with
$\epsilon=0$. The solid, dotted and dash lines depict those for $b=-0.01$, $-0.1$
and
$-0.5$,
 respectively.}
\label{fig_particle1}
\end{figure}

\begin{figure}[tbh]
\epsfxsize = 2.1in
\epsffile{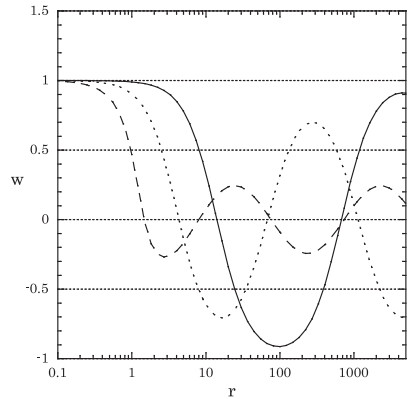}
\caption{The potential function $w(r)$ for a particle-like solution with
$\epsilon=0$.
The solid, dotted and dash lines depict those for $b=-0.01$, $-0.1$ and $-0.5$,
 respectively.}
\label{fig_particle3}
\end{figure}

For the case of $\epsilon=0$, we find the solutions which metrics are regular  in
whole spacetime and approach to Minkowski as $r\rightarrow\infty$ for $b_{\rm
min}<b<0$, where $b_{\rm min}\approx-0.635607$.  We show the
numerical result in Figs.~\ref{fig_particle1} $\sim$ \ref{fig_particle3}.

\begin{figure}[tbh]
\epsfxsize = 2.1in
\epsffile{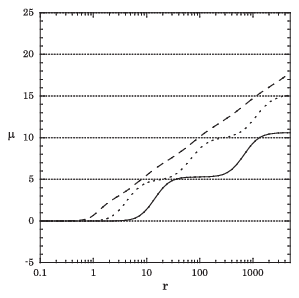}
\caption{The mass function $\mu(r)$ for a particle-like solution with
$\epsilon=0$.
The solid, dotted and dash lines depict those for $b=-0.01$, $-0.1$ and $-0.5$,
 respectively.}
\label{fig_particle2}
\end{figure}

The potential function $w$ is oscillating between $\pm 1$ and the mass function
$\mu$ is increasing without bound just as a step function.
As we show in Appendix B, there is no finite mass particle-like solution.
The mass function increases as $\ln r$ asymptotically just as the analytic
solution (\ref{nontrivialsol}).
 The period of oscillations of $w$ is the same as that of steps in
$\mu$ and it is constant in terms of $\ln r$.  This behaviour is easily understood
by solving the basic equations in the asymptotic far region ($r\rightarrow
\infty$), which  analytic forms are  
given in Appendix C.
We can check that the asymptotic solution is consistent with our numerical
solutions. The oscillations of $w$ and the periodic steps in $\mu$ are caused by
infinite number of instantons (see Appendix C).

\begin{figure}[tbh]
\epsfxsize = 2.1in
\epsffile{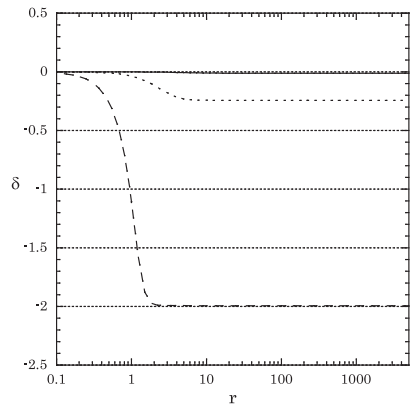}
\caption{The metric function $\delta(r)$ for a particle-like solution with
$\epsilon=1$.
The solid, dotted and dash lines depict those for $b=-0.01$, $-0.1$ and $-0.5$
 respectively.
We set $\ell=10$.}
\label{fig_particle4}
\end{figure}

\begin{figure}[tbh]
\epsfxsize = 2.1in
\epsffile{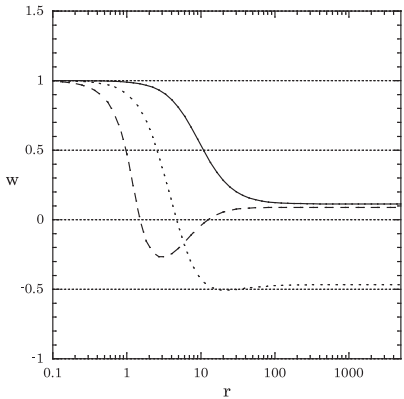}
\caption{The potential function $w(r)$ for a particle-like solution with
$\epsilon=1$.
The solid, dotted and dash lines depict those for $b=-0.01$, $-0.1$ and $-0.5$
 respectively.
We set $\ell=10$.}
\label{fig_particle6}
\end{figure}

 For the case of $\epsilon=1$, we also find a regular solution
for $b_{\rm min}<b<0$.
$b_{\rm min}$ depends on $\ell$, and decreases as $\ell$ decreases.
For example, $b_{\rm min}\approx-0.644036$ for $\ell=10$, $b_{\rm min}\approx
-1.105002$ for $\ell=1$. 
We show the numerical result in Figs.~\ref{fig_particle4} $\sim$  
\ref{fig_particle6}.

\begin{figure}[tbh]
\epsfxsize = 2.1in
\epsffile{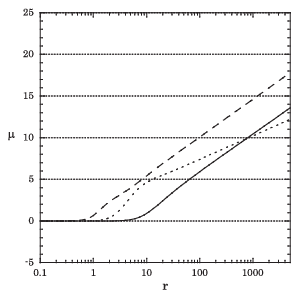}
\caption{The mass function $\mu(r)$ for a particle-like solution with
$\epsilon=1$.
The solid, dotted and dash lines depict those for $b=-0.01$, $-0.1$ and $-0.5$
respectively.
We set $\ell=10$.}
\label{fig_particle5}
\end{figure}

In this case, the potential  $w$ does not oscillate, and converge to some value
$w_{\infty}$, then the number of node is finite. The mass function 
increases monotonically as
\begin{eqnarray}
\mu\rightarrow  2(1-w_{\infty}^2)^2\ln r
\end{eqnarray}
as $r\rightarrow \infty$.
This behaviour is also understood by solving the asymptotic solution,
which is given in Appendix C.

\subsection{Black hole solution}

Next we show a non-trivial black hole solution.
To find a black hole solution, we have to impose a boundary condition at a 
horizon $r_h$. The horizon is defined by $f(r_h)=0$, which gives
\begin{eqnarray}
\mu(r_h)=r_h^2\left(1+\epsilon\frac{r_h^2}{\ell^2}\right).
\end{eqnarray}

Here we set $\delta(r_h)=0$.
The proper time of the observer at infinity (i.e. $\delta(\infty)=0$) is 
obtained  by transformation $t'=e^{-\delta(\infty)} t$.
From Eq.~(\ref{equationofw}), $w'(r_h)$ has to satisfy
\begin{eqnarray}
w'(r_h)&=&-\frac{w_h(1-w_h^2)}{r_h\left[1+2\epsilon{r_h^2}/{\ell^2}
-{(1-w_h^2)^2}/{r_h^2}\right]},
\end{eqnarray}
where $w_h=w(r_h)$.
 There is only one
free parameter $w_h$ for a given value of $r_h$. Since
Eqs.~(\ref{equationofm})-(\ref{equationofw}) are invariant under the
transformation of $w\rightarrow -w$, we can set $w_h>0$ without loss of
generality.

\begin{figure}[tbh]
\epsfxsize = 2.1in
\epsffile{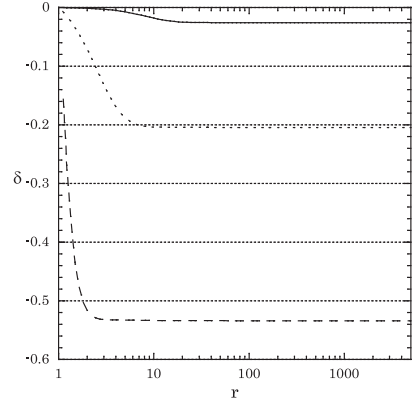}
\caption{The metric function $\delta(r)$ for a black hole solution with
$\epsilon=0$.
The solid, dotted and dash lines depict that of  $w_h=0.99$, $0.9$ and $0.5$,
 respectively.}
\label{fig_blackhole1}
\end{figure}

\begin{figure}[tbh]
\epsfxsize = 2.1in
\epsffile{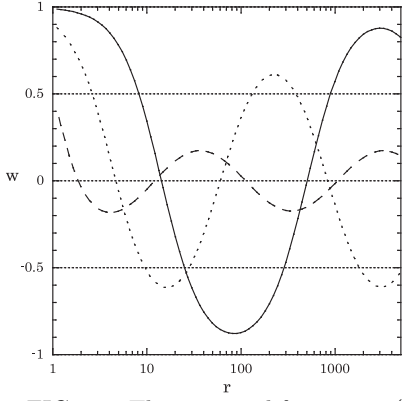}
\caption{The potential function  $w(r)$ for a black hole solution with
$\epsilon=0$.
The solid, dotted and dash lines depict that of  $w_h=0.99$, $0.9$ and $0.5$,
 respectively.}
\label{fig_blackhole3}
\end{figure}

For the solution with $w_h>1$, we find that the curvature diverges at finite 
distance. Then,  we obtain a numerical solution for $0\leq
w_h\leq 1$ for a given
$r_h$. We show the results in Figs.~\ref{fig_blackhole1} $\sim$ \ref{fig_blackhole3}.

\begin{figure}[tbh]
\epsfxsize = 2.1in
\epsffile{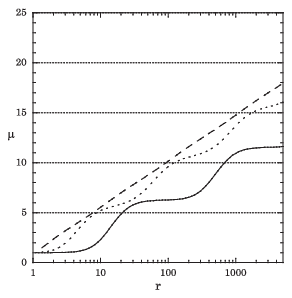}
\caption{The mass function  $\mu(r)$ for a black hole solution with
$\epsilon=0$.
The solid, dotted and dash lines depict that of   $w_h=0.99$, $0.9$ and $0.5$,
 respectively.}
\label{fig_blackhole2}
\end{figure}

Asymptotic behaviour is similar to that of the particle solution.
The potential $w$ oscillates infinitely with a constant period in terms of $\ln r$.
For any solutions with $0<w_h<1$, we find that the mass function $\mu(r)$  
diverges as $\ln r$ at large distance.

\begin{figure}[tbh]
\epsfxsize = 2.1in
\epsffile{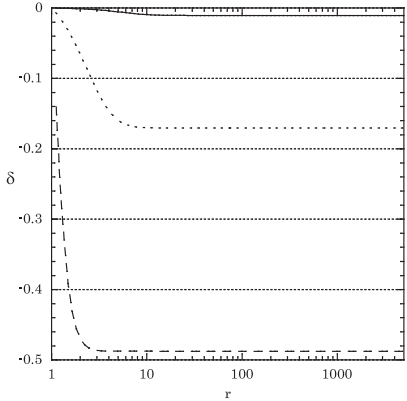}
\caption{The metric function $\delta(r)$ for a black hole solution with
$\epsilon=1$.
The solid, dotted and dash lines depict that of   $w_h=0.99$, $0.9$ and $0.5$ 
respectively.
We set $\ell=10$.}
\label{fig_blackhole4}
\end{figure}

\begin{figure}[tbh]
\epsfxsize = 2.1in
\epsffile{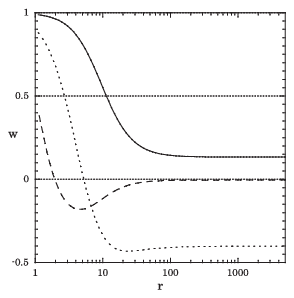}
\caption{The potential function $w(r)$ for a black hole solution with
$\epsilon=1$.
The solid, dotted and dash lines depict that of   $w_h=0.99$, $0.9$ and $0.5$ 
respectively.
We set $\ell=10$.}
\label{fig_blackhole6}
\end{figure}

We also show the case with $\epsilon =1$ in Figs.~\ref{fig_blackhole4} $\sim$ \ref{fig_blackhole6}.

\begin{figure}[tbh]
\epsfxsize = 2.1in
\epsffile{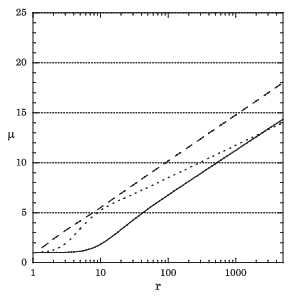}
\caption{The mass function $\mu(r)$ for a black hole solution with
$\epsilon=1$.
The solid, dotted and dash lines depict that of    $w_h=0.99$, $0.9$ and $0.5$ 
respectively.
We set $\ell=10$.}
\label{fig_blackhole5}
\end{figure}

This  also shows the similar asymptotic behaviours to a particle solution with
$\epsilon=1$.

\begin{figure}[tbh]
\epsfxsize = 2.1in
\epsffile{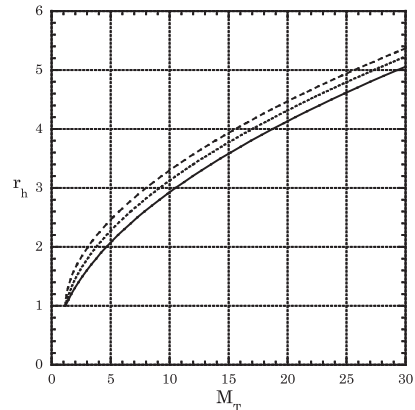}
\caption{The $r_h$-${\cal M}_T$ relation.
The solid, dotted and dash lines depict those  for
 $w_\infty=0.9$, $0.5$ and $0.0$ (analytic solution).}
\label{fig_mrh_numerical}
\end{figure}

As for the thermodynamical properties, we find the Hawking temperature  as
\begin{eqnarray}
T_{\rm BH}={e^{\delta(\infty)} \over 2\pi r_h}\left[1-{(1-w_h^2)^2\over r_h^2}
+2\epsilon{r_h^2\over \ell^2}\right] ,
\end{eqnarray}
where $\delta(\infty)$ comes from our coordinate condition, that is, we set
$\delta(r_h)=0$.
Thermodynamical mass $M_T$  is found from the first law of black hole
thermodynamics , 
$d{\cal M}_T=TdS+\Phi dQ$. 
In order to calculate $M_T$, fixing  $w_\infty=w(\infty)$,
we solve a black hole solution because a ``global charge" is
proportional to $(1-w_\infty^2)$.
The result is shown in
Fig.~\ref{fig_mrh_numerical}.

We numerically confirm that the thermodynamical mass ${\cal M}_T$ is equal to
\begin{eqnarray}
{\cal M}_T=\lim_{r\rightarrow \infty}{3\pi\over 8}\left[\mu-2(1-w^2)^2\ln 
r\right] .
\label{thermo_mass}
\end{eqnarray}

\section{Stability}
\label{sec:stability}

In this section, we analyze stability of the static solutions obtained above.
We perturb the metric and potential as
\begin{eqnarray}
\mu(r,t)&=&\mu_0(r)+ \mu_1(r)e^{i\omega t},\\
\delta(r,t)&=&\delta_0(r)+ \delta_1(r)e^{i\omega t},\\
w(r,t)&=&w_0(r)+ w_1(r)e^{i\omega t},
\end{eqnarray}
where $\mu_0(r),\delta_0(r),$ and $w_0(r)$ are those of the static  solution
obtained in previous section. Substituting them into Einstein equations and
Yang-Mills equation, we find the perturbation equations as
\begin{eqnarray}
\mu_1'&=&2 r\left[2f_0w_0'w_1'-{w_0'^2\over r^2}\mu_1
-\frac{4(1-w_0^2)w_0}{r^2}w_1\right],
\label{perturbationofm}\\
\mu_1&=&4 rf_0w_0'w_1,
\label{perturbationofm2}\\
\delta_1'&=&-\frac{4}{r}w_0'w_1',\label{perturbationofdelta}
\end{eqnarray}
and
\begin{eqnarray}
&&-\frac{1}{r^3}(rf_0e^{-\delta_0}w_0')'f_0^{-1}\mu_1 
-f_0e^{-\delta_0}w_0'\left(\frac{1}{r^2} f_0^{-1}\mu_1+\delta_1\right)'\nonumber
\\ &&+\frac{1}{r}(rf_0e^{-\delta_0}w_1')'+\frac{2}{r^2}e^{-\delta_0}(1-3w_0^2)w_1
=-\omega^2f_0^{-1}e^{\delta_0}w_1, 
\nonumber \\ 
~\label{perturbationofw}
\end{eqnarray}
where $f_0=1-\mu_0(r)/r^2+\epsilon r^2/\ell^2$.
Eq.~(\ref{perturbationofm}) is derived from 
Eq.~(\ref{perturbationofm2}) by differentiation.

We introduce a tortoise coordinate $r_*$ such that
\begin{eqnarray}
\frac{dr_*}{dr}=e^{\delta_0}f_0^{-1}
\end{eqnarray}
and define $\chi=w_1r^{\frac{1}{2}}$.
Then, by substituting Eqs.~(\ref{perturbationofm})-(\ref{perturbationofdelta}),
Eq.~(\ref{perturbationofw}) turns to be a single uncoupled equation as
\begin{eqnarray}
-\frac{d^2\chi}{dr_*^2}+V(r_*)\chi=\omega^2\chi,\label{schrodinger}
\end{eqnarray}
where
\begin{eqnarray}
V(r_*)&=&f_0e^{-\delta_0}\left\{\frac{2}{r^2}e^{-\delta_0}(3w_0^2-1)
+\frac{r^{-\frac{1}{2}}}{2}(r^{-\frac{1}{2}}f_0e^{-\delta_0})'\right.\nonumber \\
&&\left.+\frac{4}{r}\left[f_0e^{-\delta_0}w_0'^2\right]'\right\}.
\end{eqnarray}

When $V(r_*)$ is positive definite, we can prove its stability as follows:
Multiplying Eq.~(\ref{schrodinger}) by $\bar{\chi}$ and integrating from $r=r_+ 
(r_*=-\infty)$ in the case of a black hole solution or $r=0~(r* = 0)$ in the case
of a particle solution to $r=\infty (r_*=r_{*,{\rm max}} (<\infty))$,
Eq.~(\ref{schrodinger}) is written as
\begin{eqnarray}
&&-\left[\bar{\chi}\frac{d\chi}{dr*}\right]^{r=\infty}_{r=r_+ ({\rm or}~0)} 
+\int\left[\left|\frac{d\chi}{dr*}\right|^2+V(r)|\chi|^2\right]dr_*
\nonumber \\
&&~~=\omega^2\int|\chi|^2dr_*.
\label{schrodinger2}
\end{eqnarray}

We assume that $w_1\rightarrow 0$ at infinity ($r\rightarrow\infty 
(r*\rightarrow
r*_{\rm max})$). Then, $\bar{\chi}d\chi/dr*\rightarrow 0$. In the case of a black
hole, $\chi$ must be  ingoing at horizon ($r=r_+ (r*=-\infty)$). 
Since the potential $V$ vanishes at the horizon, the ingoing wave condition gives
$\chi \sim e^{i\omega r_*}$.
If we assume that 
${\rm Im} \omega< 0$, then
$\bar{\chi}d\chi/dr_*\rightarrow 0$ at horizon is obtained. 
Because $V(r)$ is
positive definite, Eq.~(\ref{schrodinger2}) 
implies that eigenvalue $\omega$ is real, that is,
${\rm Im} \omega = 0$, which contradicts with the above assumption.
Hence, we conclude that
${\rm Im} \omega\geq 0$, which means that the present system is stable.
 In the
case of a particle solution,  we should impose
$w_1=0$ at the origin ($r=0 (r_* = 0)$), then 
we find 
$\bar{\chi}d\chi/dr*\rightarrow
0$. If
$V(r)$ is positive definite, Eq.~(\ref{schrodinger2}) 
again  implies that the eigenvalue $\omega$ is real. 
Hence, in both cases, we obtain 
 that  solutions with a positive definite 
potential $V(r_*)$ are stable.

For the analytic solution (\ref{nontrivialsol}), we find
\begin{eqnarray}
V(r_*)={f_0\over 4r^4}\left[5{\cal M}-4+10\ln r-9r^2+3\epsilon {r^4\over 
l^2}\right].
\end{eqnarray}
In the case of $\epsilon=0$ or $-1$, $V(r_*)$ is negative at large distance $r$.
While, in the case of $\epsilon=1$, we  see that $V(r_*)$ is positive 
definite for sufficiently large ${\cal M}$, i.e. for
\begin{eqnarray}
{\cal M} >{1\over 5}\left(4-10\ln r_p +9 r_p^2 -3{r_p^4\over \ell^2}\right),
\end{eqnarray}
where $r_p=(9-\sqrt{81-120/\ell^2})\ell/12$ and $\ell>\sqrt{40/27}$.

\begin{figure}[tbh]
\epsfxsize = 2.1in
\epsffile{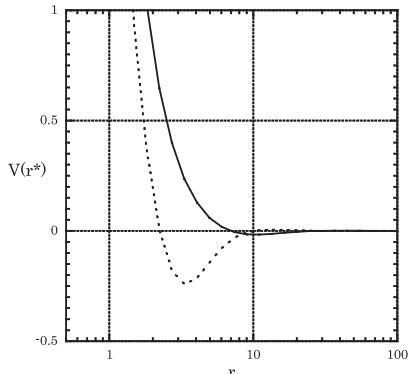}
\caption{Potential $V(r_*)$ for $\epsilon=0$.
The solid and dotted line denote those for $b=-0.01$ and $b=-0.1$.
There is a negative region ($V(r_*)<0$) for both potentials.}
\label{fig_potential1}
\end{figure}

\begin{figure}[tbh]
\epsfxsize = 2.1in
\epsffile{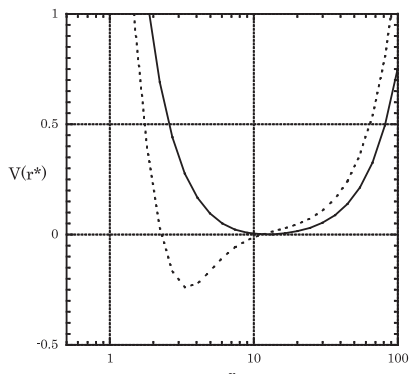}
\caption{Potential $V(r_*)$ for $\epsilon=1$.The solid and dotted line denote
those for $b=-0.01$ and $b=-0.1$. There is a negative region ($V(r_*)<0$) for
$b=-0.1$ but positive definite for
 $b=-0.01$.
We set $\ell=10$.}
\label{fig_potential2}
\end{figure}

For the numerical solutions, we also find a positive definite potential $V(r*)$
only for the  case of $\epsilon = 1$. For a particle solution, for example, 
the positive definite potential is found in the parameter range of 
$-0.010368<b<0$ for $\ell=10$, and $-0.654211<b<0$ for $\ell=1$.
We depict some typical potentials in Figs.~\ref{fig_potential1}
and \ref{fig_potential2}.

\newpage

If $V(r_*)$ is positive definite, we conclude that the system is stable, however, 
we cannot predict anything if
$V(r_*)$ is not positive definite. We have to solve
Eqs.~(\ref{perturbationofm})-(\ref{perturbationofw}) as eigenvalue problem
numerically. We leave this as a future work  and in this paper, we do not
discuss it further.

\section{Summary and Discussion}\label{sec:summaryanddiscussion}

In this paper we have studied a spherically symmetric Einstein-SU(2)-Yang-Mills 
system  in
5-dimensions. 
If we consider only ``electric part" of Yang-Mills field,
we find the 5-dimensional Reissner-Nordstrom black hole solution.
As for ``magnetic part" of Yang-Mills field,
apart from a trivial Schwarzschild (Schwarzschild-de Sitter, or Schwarzschild-anti de Sitter) solution, we find non-trivial analytic solution,
which corresponds to
a magnetically charged black hole. 
(It turns out to be just a Reissner-Nordstrom solution in 4-dimensional case).
This non-trivial solution shows that a gravitational ``mass"
is infinite and the spacetime does not 
satisfy asymptotically flat, de-Sitter, or anti-de Sitter condition, in
contrast to the case of 4-dimensions. However,
its  metric approaches either Minkowski or
de-Sitter (or anti-de Sitter) .
We also find that there is no singularity 
expect one at the origin which is covered by a horizon.
Hence we call its behaviour at infinity
``quasi-asymptotically" flat, de Sitter, or anti-de Sitter
and  we regard our solution as a localized object. 
We analyze 
the spacetime  structure and thermodynamical properties. 
They show that mass parameter 
${\cal M}$ in the solution is regarded as a thermodynamical mass, which
satisfies the first law of the black hole thermodynamics.

For the case with  zero  or negative cosmological constant,
 we also find numerically 
a particle-like solutions, which has no singularity,  and 
black hole solutions with
non-trivial structures of Yang-Mills field. 
Although, for both cases, the mass function
diverges as $\ln r$, they  satisfy 
``quasi-asymptotically" flat or anti-de Sitter conditions. 
If
$\Lambda =0$, in contrast to the case of 4-dimension, Yang-Mills
field oscillates and has infinite number of nodes.
For a negative
cosmological constant, Yang-Mills field potential settles to some constant, which is
 similar to that in 4-dimensional
case.

From stability analysis, 
we find that there is a set of stable solutions if a
cosmological constant is  negative. 
This result is very
similar to the 4-dimensional case, in which Bartnik-McKinnon solution
 and a colored black hole is unstable, while those extended to the case
with a negative cosmological constant become stable.

Since we find a stable non-singular solution in the 5-dimension,
if we apply it to a brane world scenario, we may find some interesting
effect on the brane dynamics. We will publish its analysis in a separated paper.

\acknowledgments

We would like to thank Takashi Torii for useful discussions and comments.
This work was partially supported by the Grant-in-Aid for Scientific Research 
Fund of the Ministry of Education, Science and Culture (Nos. 14047216, 14540281)
and by the Waseda University Grant for Special Research Projects.

\onecolumn

\appendix
\section{Five-dimensional Spherically Symmetric SU(2) Gauge Field}
\label{sec:appendix}

Here we calculate a generic form of spherically symmetric $SU(2)$ Yang-Mills 
field in 5-dimensional spacetime. In the case of four dimensions, Witten gave its
generic form \cite{witten}, which was called the Witten ansatz and proved by
Forg\'ac and Manton \cite{forgac}. Forg\'ac and Manton presented how to find a
generic form of spherically symmetric Yang-Mills field in arbitrary dimensions.
We just follow their method.

Suppose we have some symmetry of spacetime generated by a vector $\vec{\eta}$.
A tensor field must be invariant under an infinitesimal transformation generated 
by $\vec{\eta}$, i.e. the Lie derivative of this tensor field with respect to
$\vec{\eta}$ must vanish. However, in the case of a gauge field $A_\mu$, there is
a gauge freedom, by which we can weaken this condition such that there exists an
infinitesimal gauge transformation equivalent to a spacetime transformation, that
is
\begin{eqnarray}
\pounds_{\vec{\eta}}A_\mu&=&D_\mu W\equiv\partial_\mu W - [A_\mu, W]
\label{liederivative}
\end{eqnarray}
for some scalar field $W$ \cite{bergmann}.

Suppose that a $D$-dimensional Riemannian manifold $M$ has some spacetime 
symmetry represented by $N$-dimensional isometry group $S$. This isometry group
is generated by $N$ Killing vectors $\vec{\xi}_{(n)}~ (1\leq n\leq N)$, which
commutation relations is given by
\begin{eqnarray}
[\vec{\xi}_{(m)},\vec{\xi}_{(n)}]^\mu=f_{mnp}\xi_{(p)}^\mu,
\end{eqnarray}
where $f_{mnp}$ is a structure constant.
We assume that the orbit $X=\{ap\in M|a\in S\}$ for some point $p\in M$  is
$N'$-dimensional submanifold of $M$. Then we choose the local coordinate system as
\begin{eqnarray}
x^\mu=(x^i,y^\alpha),~~~ 1\leq i \leq D-N',~~~1\leq \alpha\leq N'
\end{eqnarray}
so that a hypersurface of $x^i={\rm constant}$ defines the orbit space $X$.
By Frobenius' theorem, the above Killing vectors are orthogonal to $X$, then 
$\xi_{(n)}^\mu=(0, \xi_{(n)}^\alpha)$ in this coordinate system.

Because the isometry group is $N$-dimensional Lie group, we can define right and 
left invariant vectors ($\vec{\xi}_{(n)}^R$ and $ \vec{\xi}_{(n)}^L$) as
\begin{eqnarray}
\pounds_{\vec{\xi}_{(n)}^R}s=sJ^{(n)} ,~~~~\pounds_{\vec{\xi}_{(n)}^L}s=-J^{(n)}s
\end{eqnarray}
for any $s\in S$, where $J^{(n)}$ is the generator of Lie group $S$ associated by 
the Killing  vector $\vec{\xi}_{(n)}$. Then both $\vec{\xi}_{(n)}^R$ and
$\vec{\xi}_{(n)}^L$ have same commutation relations as those of
$\vec{\xi}_{(n)}$. We also define those covariant vector fields
$\xi^R_{{(n)}\hat{\alpha}}, \xi^L_{{(n)}\hat{\alpha}}$ by
\begin{eqnarray}
\xi^R_{{(m)}\hat{\alpha}}\xi_{(n)}^{R\hat{\alpha}}=\delta_{mn},~~~~
\xi^L_{{(m)}\hat{\alpha}}\xi_{(n)}^{L\hat{\alpha}}=\delta_{mn}.
\end{eqnarray}

For a fixed point $q\in X$, $R=\{a\in S|aq=q\} \subset S$ is an  invariant
subgroup of $S$ with dimension $N-N'$, and the quotient group $S/R$ is
diffeomorphic to $X$. So we can adopt the same coordinates $y^\alpha$ in $X$ for
the coset $Rs\in S/R$. We take that the other coordinate components are expressed
by $y^\omega(1\leq \omega\leq N-N')$, which correspond to those of the isotropy
group $R$. If we fix the origin $s_0(y^\alpha)\in Rs(y^\alpha)$ for each coset in
smooth way, then any element s of $S$ is written uniquely with coordinates
$y^{\hat{\alpha}}=(y^\omega, y^\alpha)$, as
\begin{eqnarray}
s(y^{\hat{\alpha}})=r(y^\omega) s_0(y^\alpha)\label{localcoordinate}
\end{eqnarray}
for some $r\in R$.
In this coordinate, right invariant vector $\vec{\xi}_{(n)}^R$ is expressed  
with Killing vector $\vec{\xi}_{(n)}$ as
\begin{eqnarray}
\xi_{(n)}^{R\hat{\alpha}}=(\xi_{(n)}^{R\omega},\xi_{(n)}^{\alpha}).
\end{eqnarray}

By the above definition, we find a generic form of gauge potential $A_\mu(x^\mu)$ 
with a gauge symmetry $G$ and a spacetime symmetry $S$ as
\begin{eqnarray}
A_i(x^i,y^\alpha)=A_i(x^i),~~~~A_\alpha(x^i,y^\alpha)=\Phi_n(x^i)
\xi^L_{{(n)}\alpha}(y^\alpha)|_{y^\omega=y_0^{\omega}},\label{formofaalpha}
\end{eqnarray}
where $A_i(x^i)$ and $ \Phi_n(x^i)$  satisfy the conditions
\begin{eqnarray}
f_{mnp}\Phi_p+[\Phi_m,\Phi_n]&=&0 (^\forall m,^\forall n>N'),\nonumber \\
\partial_i\Phi_n-[A_i,\Phi_n]&=&0 (^\forall i,^\forall n>N'),\label{conditionofa}
\end{eqnarray}
and $y^\omega=y_0^{\omega}$ is a coordinate component of a unit element of the 
isotropy group $R$. Incidentally, $W_n$ in Eq.~(\ref{liederivative}) are obtained
as
\begin{eqnarray}
W_n(x^i,y^\alpha)=-\Phi_m(x^i)\xi^{R\omega}_{(n)}(y^\alpha)
\xi^L_{{(m)}\omega}(y^\alpha)|_{y^\omega=y_0^{\omega}}.\label{conditionofw}
\end{eqnarray}

Applying this formalism the 5-dimensional spherically
symmetric SU(2) gauge field, we obtain a generic form of the gauge potential 
$A_\mu(x^\mu)$. We assume the isometric group is SO(4).
In this coordinate system (\ref{metric}), the orbit $X$ is  given as $t,r={\rm
constant}$, and then $x^\mu$ is divided into $x^i=(t,r)$ and $y^\alpha=(\psi,
\theta,
\varphi)$.

The Killing vectors are given as
\begin{eqnarray}
\vec{\xi}_{(1)} &=&(0,0,-\cos\theta,\cot\psi \sin\theta,0), \nonumber\\
\vec{\xi}_{(2)} &=&(0,0,-\sin\theta \cos\varphi,- \cot\psi \cos\theta\cos\varphi,
\frac{\cot\psi \sin\varphi}{\sin\theta}), \nonumber\\
\vec{\xi}_{(3)} &=&(0,0,-\sin\theta \sin\varphi, -\cot\psi \cos\theta\sin\varphi,
-\frac{\cot\psi \cos\varphi}{\sin\theta}), \nonumber\\
\vec{\xi}_{(4)} &=&(0,0,0,-\cos\varphi,  \cot\theta \sin\varphi),\nonumber\\
\vec{\xi}_{(5)} &=&(0,0,0,-\sin\varphi,-\cot\theta \cos\varphi),\nonumber\\
\vec{\xi}_{(6)} &=&(0,0,0,0,-1),
\end{eqnarray}
and the structure constants $f_{mnp}$ are found to be
\begin{eqnarray}
f_{124}=1,~~f_{135}=1,~~f_{236}=1,~~f_{456}=1,
\end{eqnarray}
with totally antisymmetrized other components.

Next we adopt the local coordinate system which satisfies 
Eq.~(\ref{localcoordinate}) in SO(4). It is given as 4-dimensional Euler angle
$(\alpha,\beta,\chi,\psi,\theta,\varphi)$ as
\begin{eqnarray}
s(\alpha,\beta,\chi,\psi,\theta,\varphi)&=&r(\alpha,\beta,\chi)
s_0(\psi,\theta,\varphi)\nonumber\\
&=&R_{xy}(\alpha)R_{yz}(\beta)R_{xy}(\chi)R_{zu}(\psi)R_{yz}(\theta)
R_{xy}(\varphi),
\end{eqnarray}
where $R_{pq}$ denotes a rotation matrix of the $pq$-plane.
Note that $R_{xy}(\alpha) R_{yz}(\beta) R_{xy}(\chi)$ describes any element  of
an isotropy group $R$.

In this coordinate system, the right invariant vector $\vec{\xi}_n^R$ and the 
covariant left invariant vector $\vec{\xi}^L_n$ are
\begin{eqnarray}
\xi_{(1)}^{R\hat{\alpha}}&=&\left(-\frac{\sin\chi \sin\theta}{\sin\beta\sin\psi},
-\frac{\cos\chi \sin\theta}{\sin\psi},\frac{\cot\beta
\sin\chi\sin\theta}{\sin\psi},-\cos\theta,\cot\psi \sin\theta,0\right),
\nonumber\\
\xi_{(2)}^{R\hat{\alpha}}&=&\left(\frac{\cos\chi \sin\varphi  +
\sin\chi\cos\theta\cos\varphi}{\sin\beta \sin\psi},-\frac{\sin\chi \sin\varphi -
\cos\chi\cos\theta\cos\varphi}{\sin\psi}\right.,\nonumber\\ & &\left.
-\frac{\cot\beta \sin\chi \cos\theta \cos\varphi + (\cos\psi\cot\theta +
\cot\beta \cos\chi) \sin\varphi}{\sin\psi},-\sin\theta\cos\varphi,-\cot\psi
\cos\theta \cos\varphi,\frac{\cot\psi\sin\varphi}{\sin\theta}\right),  
\nonumber\\
\xi_{(3)}^{R\hat{\alpha}}&=&\left(-\frac{\cos\chi \cos\varphi  -
\sin\chi\cos\theta\sin\varphi}{\sin\beta \sin\psi},\frac{\sin\chi \cos\varphi +
\cos\chi\cos\theta\sin\varphi}{\sin\psi}\right.,\nonumber\\ &
&\left.-\frac{\cot\beta \sin\chi \cos\theta \sin\varphi - (\cos\psi\cot\theta +
\cot\beta \cos\chi)\cos\varphi}{\sin\psi},-\sin\theta\sin\varphi,-\cot\psi
\cos\theta \sin\varphi,-\frac{\cot\psi\cos\varphi}{\sin\theta}\right), 
\nonumber\\
\xi_{(4)}^{R\hat{\alpha}}&=&\left(0,0,-\frac{\sin\varphi}{\sin\theta},0,
-\cos\varphi, \cot\theta\sin\varphi\right), \nonumber\\
\xi_{(5)}^{R\hat{\alpha}}&=&\left(0,0,\frac{\cos\varphi}{\sin\theta},0,
-\sin\varphi,-\cot\theta\cos\varphi\right), \nonumber\\
\xi_{(6)}^{R\hat{\alpha}}&=&\left(0,0,0,0,0,-1\right),\\[.5em]
&&{\rm and}\nonumber\\[.5em]
\xi^L_{(1) \hat{\alpha}}&=&(0,0,0,\cos\beta, \sin\beta \cos\chi\sin\psi,\sin\beta
\sin\chi \sin\psi \sin\theta), \nonumber\\
\xi^L_{(2) \hat{\alpha}}&=&(0,0,0,-\cos\alpha \sin\beta, 
-\sin\alpha\sin\chi\sin\psi + \cos\alpha \cos\beta \cos\chi \sin\psi,  
\nonumber\\
 & & \sin\alpha \cos\chi \sin\psi \sin\theta + \cos\alpha
\cos\beta\sin\chi\sin\psi \sin\theta), \nonumber\\
\xi^L_{(3) \hat{\alpha}}&=&(0,0,0,\sin\alpha\sin\beta,-\cos\alpha\sin\chi 
\sin\psi- \sin\alpha \cos\beta \cos\chi \sin\psi,\nonumber\\ & &\cos\alpha
\cos\chi \sin\psi \sin\theta - \sin\alpha \cos\beta\sin\chi\sin\psi \sin\theta),
\nonumber\\
\xi^L_{(4) \hat{\alpha}}&=&(0,\cos\alpha,\sin\alpha\sin\beta,0,\cos\alpha\cos\chi 
\cos\psi - \sin\alpha \cos\beta \sin\chi \cos\psi,\nonumber\\ & &  \sin\alpha
\sin\beta \cos\theta + \cos\alpha \sin\chi \cos\psi\sin\theta + \sin\alpha
\cos\beta \cos\chi \cos\psi \sin\theta),
\nonumber\\
\xi^L_{(5) \hat{\alpha}}&=&(0,-\sin\alpha,\cos\alpha\sin\beta,0,-\sin\alpha 
\cos\chi\cos\psi - \cos\alpha \cos\beta \sin\chi\cos\psi,
\nonumber\\ & &
\cos\alpha \sin\beta \cos\theta - \sin\alpha \sin\chi \cos\psi\sin\theta +
\cos\alpha \cos\beta \cos\chi \cos\psi \sin\theta),\nonumber\\
\xi^L_{(6) \hat{\alpha}}&=&(1,0,\cos\beta,0, \sin\beta \sin\chi\cos\psi, 
\cos\beta \cos\theta- \sin\beta \cos\chi \cos\psi \sin\theta).
\end{eqnarray}
The equations (\ref{conditionofa}) are given as
\begin{eqnarray}
f_{mnp}\Phi_p^a + \varepsilon^{abc}\Phi_m^b\Phi_n^c&=&0\mbox{ }(a=1,2,3,m=1,
\cdots,6,n=4,5,6), \nonumber\\
\partial_i\Phi_n^a - \varepsilon^{abc}A_i^b\Phi_n^c&=&0\mbox{ }(a=1,2,3,i=
t,r,n=4,5,6).
\end{eqnarray}
This set of equations has two types of solutions; one is the ``electric"  type
and the other is the ``magnetic" one. The former type is given by
\begin{eqnarray}
A^a_t= (0,0,A_t),~~~A^a_r=(0,0,A_r), ~~~{\rm and}~~~
\Phi^a_m=0,
\end{eqnarray}
leading to the potential form as
\begin{eqnarray}
{\bf A}= \tau_3~ (A_t ~dt + A_r ~dr),
\end{eqnarray}
Using a gauge freedom, we can set $A_r=0$.
The ``electric" type of potential is now given by
\begin{eqnarray}
{\bf A}= \tau_3~A(t,r)~ dt.
\label{pot-ele}
\end{eqnarray}

While the latter type solution is given by
\begin{eqnarray}
\begin{array}{lllllllll}
A^a_t&=&(0,0,\dot{X}),&A^a_r&=&(0,0,X^\prime),& & & \\
\Phi^a_1&=&(0,0,\phi),&\Phi^a_2&=&\pm(\phi\cos X,\phi\sin X,0),&\Phi^a_3
&=&(\phi\sin X,-\phi\cos X,0),\\
\Phi^a_4&=&\pm(\sin X,-\cos X,0),&\Phi^a_5&=&-(\cos X,\sin X,0),&\Phi^a_6
&=&(0,0,\pm 1).
\end{array}
\end{eqnarray}
We then obtain a general form of $A_\mu^a$ as
\begin{eqnarray}
{\bf A} &=&\tau_3(\dot{X}dt +X^\prime dr +\phi d\psi +\cos \theta d\varphi) 
+\cos \psi \left[(\tau_1 \sin X -\tau_2 \cos X)d\theta-(\tau_1 \cos X +\tau_2\sin
X)\sin \theta d\varphi\right] \nonumber \\ &&+\phi \sin \psi \left[(\tau_1 \cos X
+\tau_2\sin X)d\theta +(\tau_1 \sin X -\tau_2 \cos X) \sin \theta
d\varphi\right].\label{formofapsi}
\end{eqnarray}
$X$ is not a dynamical variable but it is regarded as a gauge variable.
In fact, the field strength $F_{\mu\nu}^{~~a}$ is given by
\begin{eqnarray}
{\bf F}&=&\tau_3\left[ \dot{\phi} dt\wedge d\psi +\phi' dr\wedge d\psi 
-(1-\phi^2)(\sin\psi d\theta)\wedge(\sin\psi\sin\theta d\varphi)\right]\nonumber
\\ & & +(\tau_1\cos X+\tau_2\sin X)\left[ \dot{\phi}dt\wedge (\sin\psi
d\theta)+\phi'dr\wedge(\sin\psi d\theta)+(1-\phi^2)d\psi\wedge(\sin\psi\sin\theta
d\varphi)\right]\nonumber \\ & & +(\tau_1\sin X-\tau_2\cos
X)\left[\dot{\phi}dt\wedge(\sin\psi\sin\theta
d\varphi)+\phi'dr\wedge(\sin\psi\sin\theta
d\varphi)-(1-\phi^2)d\psi\wedge(\sin\psi d\theta)\right].\label{pot-mag}
\end{eqnarray}
Rotating the $\tau_1$-$\tau_2$ plane of the interior space by $-X$, the variable 
$X$ is eliminated. If we choose $X=0$, we find
\begin{eqnarray}
{\bf A} &=&\tau_3(\phi d\psi +\cos \theta d\varphi) -\cos \psi \left[\tau_2 
d\theta+\tau_1  \sin \theta d\varphi\right] \nonumber+\phi \sin \psi
\left[\tau_1  d\theta  -\tau_2\sin \theta d\varphi\right]\\ {\bf
F}&=&\tau_3\left[ \dot{\phi} dt\wedge d\psi +\phi' dr\wedge d\psi
-(1-\phi^2)(\sin\psi d\theta)\wedge(\sin\psi\sin\theta d\varphi)\right]\nonumber
\\ & & +\tau_1\left[ \dot{\phi}dt\wedge (\sin\psi d\theta)+\phi'dr\wedge(\sin\psi
d\theta)+(1-\phi^2)d\psi\wedge(\sin\psi\sin\theta d\varphi)\right]
\nonumber \\ &
& -\tau_2\left[\dot{\phi}dt\wedge(\sin\psi\sin\theta
d\varphi)+\phi'dr\wedge(\sin\psi\sin\theta
d\varphi)-(1-\phi^2)d\psi\wedge(\sin\psi d\theta)\right].
\end{eqnarray}

\twocolumn

\section{Non-existence of finite mass object ($\Lambda \leq 0$)}
\label{sec:appendixB}

Here we  show that there is no particle-like  solution with finite mass
if $\Lambda \leq 0$ ($\epsilon =0$ or 1).

Introducing new variable 
\begin{eqnarray}
z=2 \ln r
\end{eqnarray}
we rewrite the basic equations (\ref{eq_m}) and (\ref{eq_w}) with Eq. 
(\ref{eq_delta})  as
\begin{eqnarray}
&&{d\mu \over dz}=4f\left({dw\over dz}\right)^2
+\left(1-w^2\right)^2
\label{basic_eqn1}\\
&&f{d^2w\over dz^2}+\left[e^{-z}\mu+{\epsilon \over \ell^2}
e^z -e^{-z} (1-w^2)^2\right]{dw\over dz}
\nonumber \\
&&~~~+{1\over 2}w(1-w^2)=0   
\label{basic_eqn2}
\end{eqnarray}
with
\begin{eqnarray}
f=1-e^{-z}\mu+{\epsilon \over \ell^2}e^z ,
\end{eqnarray}
where the function $\delta$ is eliminated.

If we turn off gravity, that is, if we consider the Yang-Mills field
equation in the Minkowski space, we have one basic equation
\begin{eqnarray}
{d^2w\over dz^2}-
e^{-z} (1-w^2)^2{dw\over dz}
+{1\over 2}w(1-w^2)=0   .
\label{eqn_instanton}
\end{eqnarray}
This is easily integrated as
\begin{eqnarray}
{1\over 2} \left({dw \over dz}\right)^2-{1\over 8}\left(1-w^2\right)^2=E_0 ,
\label{instanton_energy}
\end{eqnarray}
where $E_0$ is an integration constant. Integrating this equation with the
boundary condition $w \rightarrow \pm 1$ as $z\rightarrow -\infty$
($r\rightarrow 0$) and $z\rightarrow \infty$ ($r\rightarrow \infty$), which
implies $E_0=0$,  we obtain the solution for
$w$ as
\begin{eqnarray}
w=\pm \tanh {z\over 2} .
\label{instanton_sol1}
\end{eqnarray}
This is exactly the same as the  Yang-Mills instanton solution in 
4-dimensional Euclidean spacetime\cite{YM_instanton}. 
If we regard 
\begin{eqnarray}
U(z)=-{1\over 8}\left(1-w^2\right)^2
\end{eqnarray}
 as a potential,  Eq. (\ref{instanton_energy})  just denotes the energy
conservation. The instanton corresponds to zero energy solution, in
which $w$ varies from $\pm 1$ to $\mp 1$ as $z= -\infty\rightarrow \infty$.

When we include the effect of gravity, whether we still have such a non-trivial 
structure or not ?  This is our question. In this appendix, we will show that
there is no   self-gravitating non-trivial solution with a finite mass energy.
To discuss it, 
we introduce the energy function $E$ by
\begin{eqnarray}
E={1\over 2}f \left({dw \over dz}\right)^2-{1\over 8}\left(1-w^2\right)^2 .
\label{instanton_E}
\end{eqnarray}
The basic equations  (\ref{basic_eqn1}) and  (\ref{basic_eqn2}) are described as 
\begin{eqnarray}
&&{dE\over dz}=-4e^{-z}\left({dw \over dz}\right)^2
\left[E+{\mu\over 8}+{\epsilon \over 8\ell^2}e^{2z}\right]
\label{eq_E}\\
&&{d\mu\over dz}=8E+2\left(1-w^2\right)^2 .
\label{eq_m1}
\end{eqnarray}

Since we are interested in  a particle-like solution, which  must be 
regular at the origin, we can expand the functions $\mu$ and $w$ as
\begin{eqnarray}
&&\mu=\mu_1 e^{z} +\mu_2 e^{2z}+\mu_3 e^{3z} +\cdots
\nonumber \\
&&w=1+w_1 e^{z} +w_2 e^{2z}+w_3 e^{3z}+\cdots, 
\label{expansion_mw}
\end{eqnarray}
as $z \rightarrow -\infty (r\rightarrow 0)$.
Inserting this form into Eqs. (\ref{basic_eqn1}) and (\ref{basic_eqn2}),
we find the expansion coefficients as 
\begin{eqnarray}
&&\mu_1 =0 ~~~ \mu_2=  4w_1^2 ~~~\mu_3 =-{4\epsilon\over 
\ell^2}w_1^2+{16\over 3}w_1^3(1+w_1) \nonumber \\
&& w_2 =-{2\epsilon\over 3\ell^2}w_1+{w_1^2\over 6}(3+8w_1)  \nonumber \\
&&w_3   ={\epsilon^2\over 2\ell^4}w_1-{\epsilon\over 8\ell^2}
w_1^2(5+24w_1)
\nonumber \\
&&
~~~~~~~~~+{w_1^3\over 4}(1+8w_1)(1+2w_1)
\label{exp_coeff1}
\end{eqnarray}
where $w_1$is a free parameter.

Putting those relations in Eqs. (\ref{instanton_E}) and (\ref{expansion_mw}),
 we find
\begin{eqnarray}
&&E=-{1\over 6}w_1^2e^{3z}\left({\epsilon \over \ell^2}
+4w_1^2\right)\\
&&\mu=4 w_1^2e^{2z} .
\end{eqnarray}
For  $\epsilon=0$ or 1,
 $E\rightarrow -0$ and $\mu \rightarrow +0$
as $ z\rightarrow -\infty$.
The r.h.s. of Eq. (\ref{eq_m1}) is positive definite because 
\begin{eqnarray}
8E+2\left(1-w^2\right)^2=4f \left({dw \over dz}\right)^2+
\left(1-w^2\right)^2  , 
\end{eqnarray}
and $f>0$ should be imposed for a particle-like solution.
Hence, the mass function $\mu$ is also positive definite.

Next, we analyze the behavior of the solution near infinity ($z\rightarrow
\infty$) . If the mass function does not diverge, we can expand
$\mu $  and $w$ as 
\begin{eqnarray}
&&\mu={\cal M}_0 +{\cal M}_1 e^{-z} +{\cal M}_2 e^{-2z}+{\cal M}_3 
e^{-3z} +\cdots \nonumber\\
&&w=-1+{\cal W}_1 e^{-z} +{\cal W}_2 e^{-2z}+{\cal W}_3 e^{-3z}
+\cdots , 
\end{eqnarray}
as $z\rightarrow \infty$.

From the basic equations, we find the relations between the expansion
coefficients as
\begin{eqnarray}
&&{\cal M}_1 =0 , ~~~ {\cal M}_2=  -4{\cal W}_1^2 \nonumber \\
&&{\cal M}_3 =
{4\over 3}{\cal W}_1^2(4{\cal W}_1-3{\cal M}_0) \nonumber \\
&& {\cal W}_2 ={2\over 3}{\cal M}_0{\cal W}_1-{1\over 2}{\cal W}_1^2
\nonumber\\
&&{\cal W}_3 ={{\cal W}_1\over 8}\left(4{\cal M}_0^2-5{\cal M}_0
{\cal W}_1+2{\cal W}_1^2\right)
\end{eqnarray}
for $\epsilon=0$ , and
\begin{eqnarray}
&&{\cal M}_1 =-4{{\cal W}_1^2\over \ell^2} ~~~ {\cal M}_2=  -
{8\over \ell^2}{\cal W}_1{\cal W}_2
 \nonumber \\
&& {\cal M}_3 =
-{2\over 3}{\cal W}_1^2\left(4{\cal M}_0-3{\cal W}_1\right) ~~~{\cal W}_2 =0 
\nonumber \\ &&{\cal W}_3 ={\ell^2\over 12}{\cal W}_1 \left(4{\cal M}_0-3
{\cal W}_1\right) 
\end{eqnarray}
for $\epsilon=1$.
Here, ${\cal M}_0$ and ${\cal W}_1$ are free parameters.

Using those relations, the energy function $E$ near infinity is evaluated as 
\begin{eqnarray}
E={1\over 6}e^{-3z}{\cal W}_1^2+\cdots \rightarrow +0
\end{eqnarray}
for $\epsilon=0$
\begin{eqnarray}
E={1\over 2\ell^2}e^{-z}{\cal W}_1^2+\cdots \rightarrow +0
\end{eqnarray}
for 
$\epsilon=1$.

Since $E \rightarrow -0$ near the origin while $E\rightarrow +0$ at infinity, 
if the solution is regular everywhere,
$E$ must vanishes at some finite point ($z_0$)
and  $dE/dz \geq 0$ there.
On the other hand, 
Eq. (\ref{eq_E}) yields $dE/dz \leq 0$ since $E(z_0)=0$ and $\mu(z_0)>0$.
As a result, we have $dE/dz(z_0)=0$.
Using Eq. (\ref{eq_E}), we then find 
$dw/dz(z_0)=0$.  $E(z_0)=0$ with this equation implies  $w(z_0)=\pm 1$.
Solving the basic equations (\ref{basic_eqn1}) and (\ref{basic_eqn2}) with the
above initial values at $z_0$ ($w(z_0)=\pm 1, dw/dz(z_0)=0, \mu(z_0)=$ positive
and finite), we find a trivial solution ($w(z)=\pm 1, \mu(z)=$ a positive
constant). We conclude that there is no non-trivial particle-like solution with
a finite  mass for $\Lambda \leq 0$.

\section{Asymptotic solution ($\Lambda \leq 0$)}
\label{sec:appendixC}

We present the asymptotic solution of the present system with $\Lambda \leq 0$.
As we proved for a particle-like solution in the previous appendix  and
numerically solved for more generic case,  the mass function $\mu$ seems to
diverge. Here we solve the basic equations with some ansatz and find 
the analytic solution in the asymptotically far region. 

First, we consider the case of
$\epsilon=0$. As our ansatz, we adopt 
\begin{eqnarray}
\mu\approx {\cal M}_L z ,
\label{ansatz}
\end{eqnarray}
which is suggested from numerical solutions and also confirmed 
from the following result.
The basic equation for the Yang-Mills field is now written as
\begin{eqnarray}
&&{d^2w\over dz^2}+e^{-z}\mu {dw\over dz}+{1\over 2}w(1-w^2)
\nonumber \\
&&
\approx {d^2w\over dz^2}+{1\over 2}w(1-w^2) =0 ,
\end{eqnarray}
 as $z\rightarrow \infty$.
We can integrate this equation  as 
\begin{eqnarray}
{1\over 2}\left({dw\over dz}\right)^2-{1\over 8}\left(1-w^2\right)^2 
=E_0 ,
\label{first_int}
\end{eqnarray}
where  $E_0$ is an integration constant and denotes the asymptotic value of the
energy.  $E_0$ must be negative, otherwise $w$ diverges as $z\rightarrow \infty$.

Rewriting  Eq. (\ref{first_int}), we find 
\begin{eqnarray}
{dw\over dz} &=&\pm {1\over 2} \sqrt{8E_0+\left(1-w^2\right)^2 }
\nonumber \\
&=&\pm {1\over 2}\sqrt{\left(w_-^2-w^2\right)^2 
\left(w_+^2-w^2\right)^2 }, 
\label{first_int2}
\end{eqnarray}
where
\begin{eqnarray}
w_\pm =\sqrt{1\pm 2 \sqrt{-2E_0}},
\end{eqnarray}
which is integrated as 
\begin{eqnarray}
w=\pm w_-{\rm sn} \left({w_+\over 2}z, k\right) ,
\end{eqnarray}
where $k=w_-/w_+$.  
$w$ is oscillating in a potential $U(z)=-{1\over 8}\left(1-w^2\right)^2 $ with a
negative energy $E_0$.

In order to check out ansatz, we also solve the mass function with the above 
solution of $w$.
The mass function $\mu$ is obtained by integration of
Eq. (\ref{eq_m1}), that is
\begin{eqnarray}
\mu&=&\int dz \left(8E_0 +2(1-w^2)^2 \right)
\nonumber \\
&=&\mu_0 z + 4w_+ w_-^2 
 \int^{w_+z/2} dx \left[(1-k^2){\rm cn}^2 (x,k) \right.\nonumber \\
&&\left.~~+k^2 {\rm cn}^4 (x,k) \right] ,
\end{eqnarray}
where $\mu_0=8E_0+2(1-w_-^2)^2=-8E_0=(1-w_-^2)^2$.
The integration of the functions ${\rm cn}^2$ and ${\rm cn}^4$
is evaluated by the elliptic functions as:
\begin{eqnarray}
&&\int  dx ~{\rm cn}^2 (x,k) ={1\over k^2}\left[ -(1-k^2)x
 \right.\nonumber \\
&&~~\left. + {E(\sin^{-1}({\rm sn}(x,k)), k)
\times (1-k^2 {\rm sn}^2(x, k))\over {\rm dn}^2(x, k)}\right]
\nonumber \\
&&\int dx  ~{\rm cn}^4 (x,k) = {1\over 3 k^2}\left[(1 - k^2) x 
\right. \nonumber \\
&&~~\left. + 
    {\rm cn}(x, k) {\rm dn}(x, k) {\rm sn}(x, k)\right] + \left[
      {2(2 k^2-1)\over 3k^4}\left(-(1-k^2) x 
\right. \right.\nonumber \\
&&~~ \left. \left. + {E(\sin^{-1}({\rm sn}(x,k)), k)\times (1-k^2  {\rm sn}^2(x,
k))\over {\rm dn}^2(x,  k)}\right)\right]
\end{eqnarray}
There functions increase with oscillations as $z\rightarrow \infty$.
When we take an average of  those functions over the period of oscillation,
the averaged values are linearly increasing as
our ansatz (\ref{ansatz}).

Dividing those functions into two parts (linear functions and oscillating
functions), we find
\begin{eqnarray}
\mu={\cal M}_L z + {8\sqrt{2}k^2\over (1+k^2)^{3/2}}D\mu \left({w_+\over
2}z,k\right) ,
\end{eqnarray}
where
\begin{eqnarray}
{\cal M}_L &=&{1\over (1+k^2)^2}\left[(1-k^2)^2
\right. \nonumber \\
&& \left. +8k^2\{(1-k^2)C_2+k^2 C_4\}\right]\\
D\mu (x,k)&=&(1-k^2)\int dx \left[{\rm cn}^2(x,k) -C_2 \right]
\nonumber \\
&&+k^2 \int dx \left[{\rm cn}^4(x,k) -C_4\right]
\end{eqnarray}
\begin{eqnarray}
C_2&=& {1\over K(k)}\int^{K(k)}_0 dx~ {\rm cn}^2(x,k) \nonumber \\
C_4&=& {1\over K(k)}\int^{K(k)}_0 dx~ {\rm cn}^4(x,k) 
\end{eqnarray}
The energy $E_0$ and the amplitude $w_-$
are found to be 
\begin{eqnarray}
E_0&=& -{1\over 8}\left({1-k^2\over 1+k^2}\right)^2\nonumber \\
w_-&=& {\sqrt{2} k \over \sqrt{1+k^2}} .
\end{eqnarray}
with  $k ~(0\leq k \leq 1)$.

The asymptotic solution is then given as 
\begin{eqnarray}
w&=&\pm w_-{\rm sn} \left({w_+\over 2}z, k\right) \\
\mu&=&{\cal M}_L z + {8\sqrt{2}k^2\over (1+k^2)^{3/2}}D\mu \left({w_+\over
2}z,k\right),
\end{eqnarray}
$D\mu$ is a periodic function
with a constant period, which 
 in $z$-coordinate is 
\begin{eqnarray}
\Delta z={8\over w_+} K(k)={4\pi\over w_+} 
F\left({1\over2},{1\over2},1,k^2\right) ,
\end{eqnarray}
where
\begin{eqnarray}
w_+= {\sqrt{2}  \over \sqrt{1+k^2}} 
\end{eqnarray}

Note that if we take a limit of $k\rightarrow 1$, $E_0 \rightarrow -0$, we
recover the instanton solution, that is, $w$ is oscillating between $\pm
1$.   The width of one instanton ($w\sim
\pm1 \rightarrow \mp1$  )  is given by $\Delta z/2$, 
which diverges in this limit.
However this oscillation is repeated infinitely when a gravitational effect is
included, that is  infinite number of instantons appear in the present system.
This is  why the mass function diverges.

For the case of $\epsilon=1~(\Lambda <0)$,
since $f\rightarrow  \ell^{-2}$,
the equation for $w$ is now 
\begin{eqnarray}
&&
{d^2 w\over dz^2} +{dw\over dz} +{\ell^2\over 2}e^{-z}w(1-w^2)
\nonumber \\
&&
\approx
{d^2 w\over dz^2} +{dw\over dz} =0 .
\end{eqnarray}
Integrating this equation, we obtain that 
\begin{eqnarray}
{\Big |}{dw\over dz}{\Big |}\propto e^{-z}, 
\end{eqnarray}
which gives  the asymptotic behaviors of $w$ as
\begin{eqnarray}
w={\cal W}_0 +{\cal W}_1  e^{-z} +\cdots .
\end{eqnarray}
The equation for $\mu$ is
\begin{eqnarray}
{d\mu\over dz} &=&{4\over \ell^2}\left({dw\over dz}\right)^2
+(1-w^2)^2
\nonumber \\
&
\approx
&
(1-{\cal W}_0^2)^2
\nonumber \\
&&~~+4{\cal W}_1\left[{{\cal W}_1\over \ell^2}-
{\cal W}_0(1-{\cal W}_0^2)\right] e^{-z} +\cdots
\end{eqnarray}
and then $\mu$ is given by 
\begin{eqnarray}
\mu={\cal M}_L z +{\cal M}_0 +{\cal M}_1 e^{-z}+\cdots , 
\end{eqnarray}
where
\begin{eqnarray}
{\cal M}_L &=& (1-{\cal W}_0^2)^2\nonumber \\
{\cal M}_1 &=& -4{\cal W}_1
\left[{1\over \ell^2}{\cal W}_1-
{\cal W}_0(1-{\cal W}_0^2)\right]
\end{eqnarray}
 The energy function $E$ is evaluated as 
\begin{eqnarray}
E&=&{1\over 2\ell^2}e^{-z} \left({dw\over dz}\right)^2
-{1\over 8} (1-w^2)^2 
\nonumber \\
&\rightarrow &
-{1\over 8} (1-{\cal W}_0^2)^2 (=E_0) .
\end{eqnarray}
Since the damping rate of the energy is given by 
\begin{eqnarray}
{dE\over dz} &=&-4e^{-z} \left({dw\over dz}\right)\left[
E+{\mu\over 8} +{1\over 8\ell^2} e^{2z}\right]
\nonumber \\
&\rightarrow &
-{1\over 2\ell^2} {\cal W}_1^2 e^{-z},
\end{eqnarray}
the energy damping ceases very soon.
As a result, the energy of the system approaches some finite value ($E_0$).
This is because
the potential term drops exponentially while the  adiabatic damping term
remains.
The solution does not oscillate because the potential term
becomes ineffective quickly.


\begin{thebibliography}{99}

\bibitem{M-theory}
P. Horava and E. Witten,
Nucl. Phys. {\bf 460}, 506 (1996)

\bibitem{brane}
J. Dai, R. G. Leigh and J. Polchinski,
Mod. Phys. Lett. A {\bf 4}, 2073 (1989)

\bibitem{brane_cosmology}
P. Binetruy, C. Deffayet and D. Langlois,
Nucl. Phys. B {\bf 565}, 269 (2000);
N. Kaloper,
Phys. Rev. D {\bf 60}, 123506 (1999);
C. Csaki, M. Graesser C. Kolda and J. Terning,
Phys. Lett. B {\bf 462}, 34 (1999);
T. Nihei,
Phys. Lett. B {\bf 465}, 81 (1999)
P. Kanti, I. I. Kogan, K. A. Olive and M. Prospelov,
Phys. Lett. B {\bf 468}, 31 (1999);
J. M. Cline, C. Grojean and G. Servant,
Phys. Rev. Lett. {\bf 83}, 4245 (1999);
P. Binetruy, C. Deffayet, U. Ellwanger and D. Langlois,
Phys. Lett. B {\bf 477}, 285 (2000);
S. Mukohyama, T. Shiromizu and K. Maeda,
Phys. Rev. D {\bf 62}, 024028 (2000)

\bibitem{Kraus}
P. Kraus
J. High Energy Phys. {\bf 12}, 011 (1999)

\bibitem{maldacena}
J. M. Maldcena,
Adv. Theor. Math. Phys. {\bf 2}, 231 (1998)

\bibitem{strominger}
A. Strominger,
J. High Energy Phys. {\bf 0110}, 034 (2001)

\bibitem{5DBH}
C. Csaki, J. Erlich and C. Grojean,
Nucl. Phys. B {\bf 604}, 312 (2001)

\bibitem{bartnik}
R. Bartnik and J. McKinnon,
Phys. Rev. Lett. {\bf 61}, 141 (1988)

\bibitem{bizon}
P. Bizon,
Phys. Rev. Lett. {\bf 64}, 2844 (1990)

\bibitem{volkov}
M. S. Volkov, N. Straumann, G. Lavrelashvili, M. Heusler and O. Brodbeck,
Phys. Rev. D {\bf 54}, 7243 (1996)

\bibitem{torii}
T. Torii, K. Maeda and T. Tachizawa,
Phys. Rev. D {\bf 52}, R4272 (1995)

\bibitem{bjoraker}
J. Bjoraker and Y. Hosotani,
Phys. Rev. Lett. {\bf 84}, 1853 (2000)

\bibitem{brodbeck}
O. Brodbeck, M. Heusler, G. Lavrelashvili, N. Straumann and M. S. Volkov,
Phys. Rev. D {\bf 54}, 7338 (1996)

\bibitem{winstanley}
E. Winstanley,
Class. Quantum Grav. {\bf 16}, 1963 (1999)

\bibitem{RS}
L. Randall and R. Sundrum,
Phys. Rev. Lett. {\bf 83}, 4690 (1999)

\bibitem{Lukas}
A. Lukas, B. A. Ovrut, K. S. Stelle and D. Waldram
Phys.Rev. D {\bf 59}, 086001 (1999)

\bibitem{hosotani}
M. Kubo, C. S. Lim and H. Yamashita,
hep-ph/0111327

\bibitem{adm}
R. Arnowitt, S. Deser and C. W. Misner,
{\it Gravitation; An Introduction to Current Research}, ed. L.  Witten (New York:
Wiley) (1962)

\bibitem{ad}
L. Abbott and S. Deser,
Nucl. Phys. B {\bf 195}, 76 (1982)

\bibitem{nakao}
K. Nakao, T. Shiromizu, K. Maeda,
Class. Quantum Grav. {\bf 11}, 2059 (1994)

\bibitem{ashtekar}
A. Ashtekar and A. Magnon,
Class. Quantum Grav. {\bf 1}, L39 (1984)

\bibitem{ashtekar2}
A. Ashtekar and S. Das,
Class. Quantum Grav. {\bf 17}, L17 (2000)

\bibitem{nucamendi}
U. Nucamendi and D. Sudarsky,
Class. Quantum Grav. {\bf 14}, 1309 (1997)

\bibitem{hawking}
S. W. Hawking,
Commun. Math. Phys. {\bf 43}, 199 (1975)

\bibitem{witten}
E. Witten,
Phys. Rev. Lett. {\bf 38}, 121 (1977)

\bibitem{forgac}
P. Forg\'ac and N. S. Manton,
Commun. Math. Phys. {\bf 72}, 15 (1980)

\bibitem{bergmann}
P. G. Bergmann and E. J. Flaherty, Jr.,
J. Math. Phys. {\bf 19}, 212 (1978)

\bibitem{YM_instanton}
P. Bizon and Z. Tabor,
Phys. Rev. D {\bf 64}, 121701 (2001)

\end{thebibliography}
\end{document}